\documentclass[aps,pra,epsfigure,twocolumn,notitlepage,superscriptaddress]{revtex4-1}
\usepackage[colorlinks=true,linkcolor=blue,urlcolor=blue,citecolor=blue,pdfusetitle]{hyperref}
\usepackage[utf8]{inputenc}
\usepackage[english]{babel}
\usepackage{amsmath}
\usepackage[caption = false]{subfig}
\usepackage{graphicx,epstopdf}
\usepackage{blindtext}
\usepackage[table,xcdraw]{xcolor}
\usepackage{lipsum}
\usepackage{amsfonts}
\usepackage{bbm}
\usepackage{amssymb}
\usepackage{enumerate}
\usepackage{color}
\usepackage{latexsym}
\usepackage{physics}
\usepackage{times,txfonts}

\newcommand{\Ccal}{\mathcal{C}}
\newcommand{\Dcal}{\mathcal{D}}
\newcommand{\Ecal}{\mathcal{E}}

\newcommand{\1}{\mathbbm{1}}

\newcommand{\acs}[1]{{\color{red}#1}}

\begin{document}
	
	\title{Exergy of passive states: Waste energy after ergotropy extraction} 
	
	\author{F. H. Kamin}
	\email{f.Hatami@uok.ac.ir}
	\affiliation{Department of Physics, University of Kurdistan, P.O.Box 66177-15175 , Sanandaj, Iran}
	
	\author{S. Salimi}
	\email{ShSalimi@uok.ac.ir}
	\affiliation{Department of Physics, University of Kurdistan, P.O.Box 66177-15175 , Sanandaj, Iran}
	
	\author{Alan C. Santos}
	\email{ac\_santos@df.ufscar.br} 
	\affiliation{Departamento de Física, Universidade Federal de São Carlos, Rodovia Washington Luís, km 235 - SP-310, 13565-905 São Carlos, SP, Brazil}
	
	\begin{abstract}
		Work extraction protocol is always a significant issue in the context of quantum batteries, in which the notion of ergotropy is used to quantify a particular amount of energy that can be extracted through unitary processes. Given the total amount of energy stored in a quantum system, quantifying wasted energy after the ergotropy extraction is a question to be considered when undesired coupling with thermal reservoirs is taken into account. In this paper, we show that some amount of energy can be lost when we extract ergotropy from a quantum system and quantified by the exergy of passive states. Through a particular example, one shows that ergotropy extraction can be done by preserving the quantum correlations of a quantum system. Our study opens the perspective for new advances in open system quantum batteries able to explore exergy stored as quantum correlations.
		
	\end{abstract}
	
	\maketitle
	
	\section{Introduction}
The dynamics of quantum systems that are in contact with an external environment, the so-called open system, arises from the interaction between the quantum degrees of freedom and the time evolution of coherence and system-environment coupling strength. Recently, efforts have been made to design the environment-mediated charging process of quantum batteries by different scenarios \cite{farina2019charger,barra2019dissipative,kamin2020non,tabesh2020environment,zakavati2020bounds,garcia2020fluctuations,pirmoradian2019aging}, which is a path to the realization of new quantum batteries. 
	
With the advent of quantum thermodynamics, studies on quantum devices able to use quantum advantages to store and extract useful energy from physical systems~\cite{Santos:21b,PRL_Andolina,quach2020using,kamin2020entanglement,Alexia:20,Santos:20c}, called quantum batteries (QBs), have allowed to define new physical quantities. For example, the maximal work that can be extracted from a quantum system by unitary operations is called ergotropy~\cite{Allahverdyan:04}. However, when we take into account non-unitary effects on the QB due to the coupling of the system with external thermal baths~\cite{farina2019charger,barra2019dissipative}, the battery performance leads to ask how much energy is lost due to thermal effects on the system. In particular, there are situations in which ergotropy can be stored~\cite{Baris:20}, but some part of the total energy available in the system is not stored as ergotropy and, consequently, part of the total energy cannot be extracted from cyclic unitary transformations. Under this point of view, it is worth considering the amount of residual energy that cannot be extracted as useful work from open quantum batteries by unitary processes.

In this direction, in this paper we explore the amount of energy lost due to the constraint of unitary processes for work extraction. To this end, we show that the system exergy, the maximum amount of work extracted from the system bringing it into the equilibrium with a thermal bath, can be decomposed into two quantities: ergotropy and residual energy. Such residual energy cannot be extracted through a unitary process, then it refers to the non-optimal performance of a cyclic thermodynamics process for work extraction from QBs. In a general way, we show that the waste energy is quantified by the amount of exergy of the passive state (a state in which no energy can be extracted as ergotropy). We then apply our discussion to Werner states, where we study how entanglement and discord are associated with ergotropy and exergy of Werner passive states.
	
\section{Non-extractable energy by unitary processes}
	
	\subsection{Ergotropy and Exergy}
	We can quantify the extractable energy from a quantum system using the definition of ergotropy (for unitary process) and exergy (for non-unitary process). This is motivated by the definition of the amount of energy that is accessible due to the presence of the thermal bath. So, let us consider the following scenario: we have a quantum system of interest, the system is in an arbitrary non-equilibrium state $\rho_{0}$ with a Hamiltonian $H$. One may regard only unitary evolution, where the system evolves through a cyclic process. In this protocol, the Hamiltonian of the system is the same at beginning and at the end of the process, so that ergotropy is the maximum work that can be extracted from the battery under such a process as the following form~\cite{Allahverdyan:04} 
	\begin{align}\label{1}
	\mathcal{E}_{S}&=\tr(H\rho _{0})-\min _{U\in\mathcal{U}}\tr(H U\rho _{0}U^{\dagger})=\tr(H\rho_{0}) - \tr(H\varrho_{\rho_{0}}),
	\end{align}
	where $\mathcal{U}$ covers the set of unitary operations derived from Hamiltonian $H$, and $\varrho_{\rho_{0}}$ is the called passive state~\cite{Allahverdyan:04}. 
	In addition, in situations where the system-reservoir interaction is taken into account, energy can be lost by thermalization. To take into account such interaction, let us consider the thermalization process of the system, initially in state $\rho_{0}$, with a thermal bath at inverse temperature $\beta$. It is known that during the thermalization process an amount of energy is exchanged between the system and the reservoir given by the variation of the free energy~\cite{vedral2002role}
	\begin{align}
	\Sigma^{\rho_{0}\rightarrow\rho_{\beta}} = \mathcal{F}(\rho_{0})- \mathcal{F}(\rho_{\beta}) , \label{Sigma}
	\end{align}
	where $\mathcal{F}(\rho_{0})\!=\!\tr\{H\rho_{0}\}-\beta^{-1}S(\rho_{0})$ is the nonequilibrium free energy of the battery at the initial stage, $\mathcal{F}(\rho_{\beta})$ being the free energy of the final equilibrium state $\rho_{\beta}=e^{-\beta H}/Z$, with the von Neumann entropy and the partition function given by $S(\rho)\!=\!-\tr{\rho\ln\rho}$ and $Z=\mathrm{tr}\{e^{-\beta H}\}$, respectively. Since the quantity $\Sigma^{\rho_{0}\rightarrow\rho_{\beta}}$ is the amount of extractable work from a system through a process that brings the system into the equilibrium with a thermal reservoir, it is worth mentioning here that we can call it \textit{exergy}. We use it in analogy with the thermodynamics classical definition of \textit{exergy}, as defined by Zoran Rant~\cite{Rant:56}, from Greek `ex'  [$\varepsilon\xi$] and `ergon' [$\varepsilon\rho\gamma o\nu$].
	
	\begin{figure}[t!]
		\includegraphics[scale=1.0]{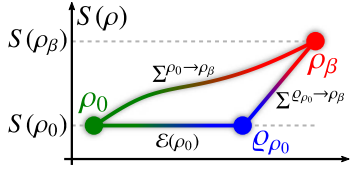}
		\caption{Schematic representation of the three states addressed in our discussion. The ergotropy extraction process, which does not change the system entropy, is followed by the thermalization process that brings the system into equilibrium with the thermal reservoir at temperature $\beta$.}\label{Fig-Scheme}
	\end{figure}
	
	\subsection{Residual energy after ergotropy extraction}
	
	It is known that after the ergotropy extraction, the internal energy of the system is not zero~\cite{Allahverdyan:04,Moraes:21}, so that a residual amount of energy is yet available in the system. In order to quantify the amount of extractable energy after ergotropy extraction, we consider the dynamics as depicted in Fig.~\ref{Fig-Scheme}. The available work in an initial state $\rho_{0}$ can be extracted in two ways: (i) a unitary process which brings the system into a passive state $\varrho_{\rho_{0}}$, so that the extractable work is quantified by ergotropy~\cite{Allahverdyan:04}, and (ii) a non-unitary process that leads the system to the thermal equilibrium state $\rho_{\beta}$, quantified by the variation of the free energy~\cite{vedral2002role}. As sketched in Fig.~\ref{Fig-Scheme}, it is possible to identify some amount of energy $\Sigma_{\text{ex}}$ so that the exergy $\Sigma(\rho_{0}\rightarrow\rho_{\beta})$ extracted in the process $\rho_{0}\rightarrow\rho_{\beta}$, and the extractable ergotropy $\mathcal{E}$ in the process $\rho_{0}\rightarrow\varrho_{\rho_{0}}$, satisfy the balance equation  $\Sigma(\rho_{0}\rightarrow\rho_{\beta}) = \mathcal{E} + \Sigma_{\text{ex}}$, where $\Sigma_{\text{ex}}$ is an available work that cannot be extracted by unitary processes. In fact, consider Eq.~\eqref{Sigma} as
	\begin{align}
	\Sigma^{\rho_{0}\rightarrow\rho_{\beta}} = \tr\{H\rho_{0}\}-\beta^{-1}S(\rho_{0}) - [\tr\{H\rho_{\beta}\}-\beta^{-1}S(\rho_{\beta})] \acs{,}
	\end{align}
	where we can assume the sum-zero equation given by $\tr\{H^{(0)}\varrho_{\rho_{0}}\} - \tr\{H^{(0)}\varrho_{\rho_{0}}\}$ in order to write
	\begin{align}
	\Sigma^{\rho_{0}\rightarrow\rho_{\beta}} &= \tr\{H\rho_{0}\} - \tr\{H\varrho_{\rho_{0}}\} - [\tr\{H\rho_{\beta}\}-\beta^{-1}S(\rho_{\beta})] \nonumber\\ &+ \tr\{H\varrho_{\rho_{0}}\} - \beta^{-1}S(\rho_{0}) .
	\end{align}
	
	Now, notice that the first two terms are associated with the ergotropy of the initial state $\rho_{0}$, so that
	\begin{align}\label{abo}
	\Sigma^{\rho_{0}\rightarrow\rho_{\beta}} &= \mathcal{E}(\rho _{0}) - \left[\tr\{H\rho_{\beta}\}-\beta^{-1}S(\rho_{\beta})\right] \nonumber \\ &+ \tr\{H\varrho_{\rho_{0}}\} - \beta^{-1}S(\rho_{0}) .
	\end{align}
	
	In addition, since the process that brings the system from state $\rho_{0}$ to the passive state $\varrho_{\rho_{0}}$ is unitary, we also can write $S(\rho_{0})\!=\!S(\varrho_{\rho_{0}})$, and Eq.~\eqref{abo} becomes
	\begin{align}\label{EX}
	\Sigma^{\rho_{0}\rightarrow\rho_{\beta}} &= \mathcal{E}(\rho _{0}) + \left[\mathcal{F}(\varrho_{\rho_{0}})- \mathcal{F}(\rho_{\beta})\right] .
	\end{align}
	
	In conclusion, as sketched in Fig.~\ref{Fig-Scheme}, the additional amount of energy $\Sigma_{\text{ex}}$ is given by the free energy of the passive state associated to $\rho_{0}$, which is equivalent to exergy stored in the passive state $\varrho_{\rho_{0}}$, mathematically
	\begin{align}
	\Sigma_{\text{ex}} = \Sigma^{\varrho_{\rho_{0}}\rightarrow\rho_{\beta}} = \mathcal{F}(\varrho_{\rho_{0}})- \mathcal{F}(\rho_{\beta}) \label{Exergy}.
	\end{align}
	
	This result can be understood in two different ways. The first interpretation refers to the uniqueness of an energetically efficient initial state for store ergotropy. In fact, given a system that interacts with a reservoir, the initial ergotropy is stored in a non-pure state $\rho_0$. Then, for a short time interval, we can drive the system in order to extract ergotropy through the optimal unitary operation $U_\text{opt}$. By adequately choosing the initial state so that $U_\text{opt}\rho_{0}U_\text{opt}^{\dagger}\rightarrow\rho_{\beta}$, it is possible to see that $\Sigma^{\varrho_{\rho_{0}}\rightarrow\rho_{\beta}}\!=\!0$, leading to $\Sigma^{\rho_{0}\rightarrow\rho_{\beta}} \!=\!\mathcal{E}(\rho _{0})$. In conclusion, all available energy of the system can be extracted as ergotropy. So, given the uniqueness of the thermal state $\rho_{\beta}$, for the optimal unitary operation $U_\text{opt}$ we have the uniqueness of $\rho_{0}$. The second case refers to the way to efficiently extract energy. In addition, given that the quantity $\Sigma^{\varrho_{\rho_{0}}\rightarrow\rho_{\beta}}$ cannot be a negative number, this means that energy lost during the ergotropy extraction is expected as a natural process due to the entropy production. This second case can be understood as an immediate application of the second law of thermodynamics to quantum batteries. In fact, for $\Sigma^{\varrho_{\rho_{0}}\rightarrow\rho_{\beta}}\!>\!0$, the Eq.~\eqref{Exergy} gives
	\begin{align}
	\Delta S_{\text{ex}} > \beta \left( \tr\{H\rho_{\beta}\} - \tr\{H\varrho_{\rho_{0}}\} \right) ,
	\end{align}
where $\Delta S_{\text{ex}}=S(\rho_{\beta})-S(\varrho_{\rho_{0}})$ is the entropy production required to extract the exergy stored in the system. Then, given that during the thermalization process the heat exchanged between the system and reservoir is given by the internal energy variation of the system~\cite{Alicki:79}, we conclude that $\Delta S_{\text{ex}}\!>\!\beta Q$.

	\section{Residual energy as quantum correlations}
	
	In this section, we show that quantum correlations after ergotropy extraction can preserve a non-zero amount of energy. To this end, we consider a two-spin Werner state given by
	\begin{align}
	\rho_{\text{w}} = \frac{1-\varepsilon}{4} \1 + \varepsilon \ket{\beta}\bra{\beta} , \label{Eq-Werner}
	\end{align}
with the Bell state $\ket{\beta}\!=\!(\ket{\uparrow\uparrow}+\ket{\downarrow\downarrow})/\sqrt{2}$. The above state is adequate for our study because we can adequately choose $\varepsilon$ to control the amount of quantum correlation of $\rho_{\text{w}}$. In particular, here we consider concurrence~\cite{wootters1998entanglement} as the measurement of entanglement, and quantum discord~\cite{PhysRevLett.88.017901,ciccarello2014toward} as correlations beyond entanglement. The concurrence is given by $\Ccal(\rho)\!=\!\mathrm{max} \{0 , \lambda_{1} -  \lambda_{2} -  \lambda_{3} - \lambda_{4} \}$, where $\lambda_{n}$ are the eigenvalues of the operator $\sqrt{\rho^{1/2}\tilde{\rho}\rho^{1/2}}$ in decreasing order, with $\tilde{\rho}=(\sigma_{y}\otimes\sigma_{y})\rho^{\ast}(\sigma_{y}\otimes\sigma_{y})$, being the complex conjugate $\rho^{\ast}$ of $\rho$ taken in the basis $\ket{0}$ and $\ket{1}$ of the system. Quantum discord is computed from the one-sided trace distance discord (TDD) $\Dcal(\rho)\!=\!D^{(\rightarrow)}(\rho)$ for $X$-states as
	\begin{align}\label{Eq-Discord}
	\Dcal(\rho) = \frac{1}{2}\sqrt{\frac{\gamma^{2}_{1} \max \{ \gamma_{3}^{2}, \gamma_{2}^{2} + x^2\} - \gamma^{2}_{2} \max \{ \gamma_{3}^{2}, \gamma_{1}^{2}\}}{\max \{ \gamma_{3}^{2}, \gamma_{2}^{2} + x^2\} - \min \{ \gamma_{3}^{2}, \gamma_{1}^{2}\} + \gamma^{2}_{1} - \gamma^{2}_{2}}} ,
	\end{align}
	where $\gamma_{1}\!=\!2(\rho_{32}+\rho_{41})$, $\gamma_{2}\!=\!2(\rho_{32}-\rho_{41}\textcolor{red}{)}$, $\gamma_{3}\!=\!1-2(\rho_{22}+\rho_{33})$, and $x\!=\!2 (\rho_{11}+\rho_{22})- 1$. Then, for the state in Eq.~\eqref{Eq-Werner}, one gets
	\begin{align}
	\Dcal(\rho_{\text{w}}) = \frac{\varepsilon}{2} , ~~ \Ccal(\rho_{\text{w}}) = \max \left[ 0 , \frac{3\varepsilon -1}{2} \right] . \label{Eq-Correlations}
	\end{align}
	
	The discord considered here is motivated by recent results shown in Ref.~\cite{Cruz:21}, where a room-temperature quantum battery has been proposed by storing ergotropy as quantum discord given in Eq.~\eqref{Eq-Discord}. We also consider the quantum discord as originally defined by Olliver and Zurek~\cite{PhysRevLett.88.017901} (see Appendix~\ref{Ap-Discord}), since an analytical expression of the discord for X-states has been proposed in Ref.~\cite{PhysRevA.84.042313}. As we shall see, both quantum discord and entanglement of the passive state depend on the Hamiltonian of the system. In general, the passive state of a system depends on the reference Hamiltonian used to set the system ergotropy. In order to illustrate this situation, we consider here two different scenarios: the Ising and Heisenberg models. As we shall see, while the energy eigenstates of the Ising Hamiltonian do not present entanglement, quantum correlations are observed for the energy eigenstates of the Heisenberg Hamiltonian.
	
	\subsection{Ising Hamiltonian}
	First, we focus on the Ising model, whose reference Hamiltonian reads as
		\begin{align}
		H_{Is}=H_{0}+H^{\mathrm{Is}}_{int},
		\end{align}
		with $H_{0}=\sum_{i=1,2}\hbar\omega\sigma^{i}_{z}$ and $H^{\mathrm{Is}}_{int}=J\hbar\sigma^{1}_{z}\sigma^{2}_{z}$ being the free and interaction Hamiltonians of the coupled qubit system, respectively. Where $\sigma^{i}_{z}~(i=1,2)$ belongs to the set of standard Pauli matrices $\sigma_{j}$ with $j\in\lbrace x,~y,~z\rbrace$ for each subsystem and, $J\!\geq\!0$ (by definition) is the strength of two-body interaction. The eigenenergies can be found by the direct diagonalization of the reference Hamiltonian $H_{Is}$ in following set 
		\begin{align}
		\text{spectrum} = \left\{ -J\hbar,~ -J\hbar,~(J-2\omega)\hbar,~(J+2\omega)\hbar  \right\} ,
		\end{align}
		in which the ordering depends on the values of $J$ and $\omega$, associated with eigenstates $\ket{\downarrow\uparrow},~\ket{\uparrow\downarrow},~\ket{\downarrow\downarrow},~\ket{\uparrow\uparrow}$, respectively. Here, $\ket{\downarrow}$ is the ground state and $\ket{\uparrow}$ is the excited state of a single qubit. Consequently, since we can have crossing levels in the transition from $J-2\omega\!\leq\!-J$ to $J-2\omega\!>\!-J$, we can be considered two feasible situations $\omega\!\geq\! J$ and $\omega\!<\! J$, to achieve an increasing order of energy levels.
		In the former case, one can obtain $\mathcal{E}(\rho_{\mathrm{w}})=2\varepsilon\omega\hbar$, where the passive state is given by
		\begin{align}
		\varrho_{\rho_{\text{w}}}^{\text{Is}}(\omega\!\geq\!J) = \frac{1-\varepsilon}{4} \1 + \varepsilon \ket{\downarrow\downarrow}\bra{\downarrow\downarrow} ,
		\end{align}
		and for the latter, we have $\mathcal{E}(\rho_{\mathrm{w}})=2\varepsilon J\hbar$ with the corresponding passive state 
		\begin{align}
		\varrho_{\rho_{\text{w}}}^{\text{Is}}(\omega\!<\!J) = \frac{1-\varepsilon}{4} \1 + \varepsilon \ket{\downarrow\uparrow}\bra{\downarrow\uparrow} .
		\end{align}
		Therefore, the maximum charge of the battery fulfill the following form
		\begin{align}\label{ergo}
		\Ecal^{\text{Is}}(\rho_{\text{w}}) = \Ecal^{\text{Is}}_{0} \varepsilon , ~~ \Ecal^{\text{Is}}_{0} = \left\{ \begin{matrix}
		2 \omega \hbar & \omega \geq J
		\\
		2 J\hbar  & \omega < J
		\end{matrix} \right. ,
		\end{align}
		for any desired value $\varepsilon$. By comparing Eqs. \eqref{Eq-Correlations} and \eqref{ergo}, we see that the ergotropy can be written in terms of the quantum discord as $\Ecal^{\text{Is}}(\rho_{\text{w}}) = 2 \Dcal(\rho_{\text{w}}) \Ecal^{\text{Is}}_{0}$. In addition, we can see that $\Ccal(\rho_{\text{w}})\!=\!0$ and $\Ecal(\rho_{\text{w}})\!\neq\!0$ for $0\!\leq\!\varepsilon\!\leq\!1/3$. This proves that, in general, entanglement is not an accessible resource for maximum work extraction compared to quantum discord in such a model. On the other hand, it is straightforward to investigate that $\Ccal(\varrho_{\rho_{\text{w}}}^{\text{Is}})$ and $\Dcal(\varrho_{\rho_{\text{w}}}^{\text{Is}})$ are zero for both $\omega\!<\!J$ and $\omega\!\geq\!J$. So it can be argued that ergotropy is entirely stored in quantum discord.
		
According to Eq.~\eqref{EX}, after the ergotropy extraction procedure, we have an amount of energy that cannot be extracted through a coherent interaction with an external field, which leads to a unitary process. Such residual amount of energy can be obtained for the Ising model as follow
\begin{align}
		\Sigma^{\varrho_{\rho_{0}}\rightarrow\rho_{\beta}} = \Sigma_{0} \varepsilon + \beta^{-1}C_{\text{Is}} , ~~ \Sigma_{0} = \left\{ \begin{matrix}
		(J-2 \omega)\hbar & \omega \geq J
		\\
		-J\hbar  & \omega < J
		\end{matrix} \right. ,
		\end{align}
where $C_{\text{Is}}\!=\!\text{ln}(Z_{\text{Is}})-S(\varrho_{\rho_{\text{w}}}^{\text{Is}})$, with $Z_{\text{Is}}\!=\!2(e^{\beta J \hbar}+e^{-\beta J\hbar}\cosh[2\beta\omega \hbar])$ the partition function of thermal state $\rho_{\beta}$ and $S(\varrho_{\rho_{\text{w}}}^{\text{Is}})$ the entropy of the passive state given by
\begin{align}
S(\varrho_{\rho_{\text{w}}}^{\text{Is}})=2\ln 2- \left[\frac{3(1-\varepsilon)}{4}\ln(1-\varepsilon)+\frac{(1+3\varepsilon)}{4}\ln(1+3\varepsilon)\right] . \label{Eq-EntIsing}
\end{align}			

It is important to mention that, besides the exergy depends on the parameter $\varepsilon$, it is not associated with any amount of correlation, since the passive state $\varrho_{\rho_{\text{w}}}^{\text{Is}}$, does not present any quantum correlation. Then, exergy of passive states for the Ising Hamiltonian is not stored as correlations.
	
	\subsection{Heisenberg Hamiltonian}
	
	The Heisenberg Hamiltonian describes a system where the interaction part is given by
	\begin{align}
	H_{\text{int}}^{\text{H}} = \frac{J\hbar}{\sqrt{2}} \left( \sigma_{x}^{1} \sigma_{x}^{2} + \sigma_{y}^{1} \sigma_{y}^{2} \right) ,
	\end{align}
	where factor $\sqrt{2}$ is considered here to give a fair comparison with the Ising model, so that $||H_{\text{int}}^{\text{H}}||\!=\!||H_{\text{int}}^{\text{H}}||$, being $||A||\!=\!(\text{tr} ( A A^{\dagger} ) )^{1/2}$ the Hilbert-Schmidt norm of the operator $A$. This quantity allows for quantifying the thermodynamic cost of implementing the dynamics driven by an arbitrary Hamiltonian~\cite{Deffner:21}. By computing the spectrum of the reference Hamiltonian $H_{\text{H}}\!=\! H_{0} + H_{\text{int}}^{\text{H}}$, we find
	\begin{align}
	\text{spectrum} = \left\{ -\sqrt{2} J\hbar, -2 \omega\hbar, \sqrt{2} J\hbar, 2 \omega \hbar \right\} ,
	\end{align}
	with eigenstates $(\ket{\downarrow\uparrow}-\ket{\uparrow\downarrow})/\sqrt{2}$, $\ket{\downarrow\downarrow}$, $(\ket{\downarrow\uparrow}+\ket{\uparrow\downarrow})/\sqrt{2}$ and $\ket{\uparrow\uparrow}$, respectively. We stress that, for this case, whenever we have $|J|\!>\!\!\sqrt{2}|\omega|$, the ground state of the system is a maximally entangled state. After a detailed analysis, it is possible to show that the ergotropy can be adequately written as
	\begin{align}\label{ergoH}
	\Ecal^{\text{H}}(\rho_{\text{w}}) = \Ecal^{\text{H}}_{0} \varepsilon , ~~ \Ecal^{\text{H}}_{0} = \left\{ \begin{matrix}
	2 \omega \hbar & \omega \geq J/\sqrt{2}
	\\
	\sqrt{2} J\hbar  & \omega < J/\sqrt{2}
	\end{matrix} \right. ,
	\end{align}
	
	Similar to the Ising case, given an arbitrary value for $\varepsilon$, it is not possible to write $\Ecal(\rho_{\text{w}})\!\propto\!\Ccal(\rho_{\text{w}})$, however, we can write $\Ecal^{\text{H}}(\rho_{\text{w}})\!=\!2 \Dcal(\rho_{\text{w}}) \Ecal^{\text{H}}_{0}$. Therefore, we conclude that the ergotropy stored in the Werner state is not stored as entanglement, but ergotropy is stored quantum discord. Then, given recent results on discord-based ergotropy~\cite{Cruz:21}, we understand that entanglement is not the main quantum resource of quantum batteries in general and confirms the results presented in \cite{kamin2020entanglement}. Then, after ergotropy extraction, the passive state for the Heisenberg Hamiltonian reads as
	\begin{align}
	\varrho_{\rho_{\text{w}}}^{\text{H}}(\omega\!\geq\!J/\sqrt{2}) = \frac{1-\varepsilon}{4} \1 + \varepsilon \ket{\downarrow\downarrow}\bra{\downarrow\downarrow} ,
	\end{align}
	for the regime where $\omega\!\geq\!J/\sqrt{2}$. In this case, we can see the amount of energy $\Ecal(\rho_{\text{w}})\!=\!2\varepsilon\hbar\omega$ is extracted from the system so that the passive state has no amount of correlation. In fact, from concurrence and discord we find $\Ccal(\varrho_{\rho_{\text{w}}}^{\text{H}})\!=\!\Dcal(\varrho_{\rho_{\text{w}}}^{\text{H}})\!=\!0$. Therefore, the maximum work extraction comes with fully ``correlation extraction". 
	
	On the other hand, by considering the case with $\omega\!<\!J/\sqrt{2}$ we get the passive state
	\begin{align}
	\varrho_{\rho_{\text{w}}}^{\text{H}}(\omega\!<\!J/\sqrt{2}) = \frac{1-\varepsilon}{4} \1 + \varepsilon \ket{\beta_{-}}\bra{\beta_{-}} ,
	\end{align}
	where $\ket{\beta_{-}}\!=\!(\ket{\uparrow\downarrow}-\ket{\downarrow\uparrow})/\sqrt{2}$ is the singlet-state of two-qubit, one of the Bell states. It is worth highlighting this case for a particular motivation: the ergotropy extraction can be done without changing the amount of correlation in the system. To the best of our knowledge, such a result was not observed so far and it has an implication different from our previous discussion. Because the states $\varrho_{\rho_{\text{w}}}^{\text{H}}(\omega\!<\!J/\sqrt{2})$ and $\rho_{\text{w}}$ have the same amount of correlation, since we can obtain $\rho_{\text{w}}$ up to local rotations in $\varrho_{\rho_{\text{w}}}^{\text{H}}(\omega\!<\!J/\sqrt{2})$, we understand that the ergotropy of the system with Heisenberg Hamiltonian, where $\omega\!<\!J/\sqrt{2}$, is not stored as correlations. It means that by increasing the interaction strength of the system, the ergotropy of Werner states can be extracted from the system without destroying correlations in the system, leading then to a Werner passive state.
	
According to the final interpretations of the previous section, it turns out that the passive state exergy can be a physical justification for such an event. To clarify this idea underlying the waste energy, we find the exergy of passive state in the following form
\begin{align}
\Sigma^{\varrho_{\rho_{0}}\rightarrow\rho_{\beta}} = \Sigma_{0} \varepsilon + \beta^{-1}C_{\text{H}} , ~~ \Sigma_{0} = \left\{ \begin{matrix}
-2 \omega\hbar & \omega \geq \frac{J}{\sqrt{2}}
\\
-\sqrt{2}J\hbar  & \omega < \frac{J}{\sqrt{2}}
\end{matrix} \right. ,
\end{align}
with $C_{\text{H}}\!=\!\ln(Z_{\text{H}})-S(\varrho_{\rho_{\text{w}}}^{\text{H}})$, with $Z_{\text{H}}\!=\!2(\cosh[\sqrt{2}\beta J\hbar]+\cosh[2\beta\omega\hbar])$ and $S(\varrho_{\rho_{\text{w}}}^{\text{H}})$ given by Eq.~\eqref{Eq-EntIsing}, since the passive state $\varrho_{\rho_{\text{w}}}^{\text{H}}$ can be obtained from $\varrho_{\rho_{\text{w}}}^{\text{Is}}$ through unitary operations for any $J$ and $\omega$. As the main result, it is possible to see that part of the exergy is stored as discord, since we can rewrite the above equation as $\Sigma^{\varrho_{\rho_{0}}\rightarrow\rho_{\beta}} = 2\Sigma_{0}\Dcal(\rho_{\text{w}}) + \beta^{-1}C_{\text{H}}$, where we have used $\varepsilon\!=\!2\Dcal(\rho_{\text{w}})$ obtained from Eq.~\eqref{Eq-Correlations}.

As shown in Appendix~\ref{Ap-Discord}, it is worth mentioning that a different quantifier of discord can be considered here. For example, we can use the original proposal, by Olliver and Zurek~\cite{PhysRevLett.88.017901}, of quantum discord and explore the analytical results obtained in Ref.~\cite{PhysRevA.84.042313}. As a result, we stress that the trace distance discord is more adequate to describe the amount of ergotropy stored in the system as quantum correlations of the system.

\section{conclusion}

This paper dealt with the loss of energy from a (cyclic) unitary work extraction from a quantum system, where such an amount of energy is identified as the exergy of the quantum passive state. We showed that, in general, in a real scenario the ergotropy leads to loss of energy due to the limitation of the unitary process. In addition, given the system-bath interaction that leads to the thermalization process, we discussed the existence and uniqueness of an optimal passive state for ergotropy and exergy extraction. From the point of view of the second thermodynamics law, we explain our main result as a natural consequence of the entropy production of the thermalization process for exergy extraction. As an application of our results, it is possible to identify a family of ergotropy and exergy extraction where the total amount of quantum correlations (as quantified by the quantum discord) of the system is conserved. Then, it implies that the exergy of a quantum passive state can be stored as quantum correlations. Since exergy is the amount of energy extractable through a thermalization process, a new prospect is opening up for exploring protocols of operational open quantum batteries.

\begin{acknowledgements}
This work has been supported by the University of Kurdistan. F. H. Kamin and S. Salimi thank Vice Chancellorship of Research and Technology, University of Kurdistan. A.C.S. acknowledges the financial support of the São Paulo Research Foundation (FAPESP) (Grant No. 2019/22685-1).
\end{acknowledgements}

\appendix
	

\section{Olliver-Zurek quantum discord} \label{Ap-Discord}

For sake of completeness, we compute here the quantum discord as originally proposed by Olliver and Zurek~\cite{PhysRevLett.88.017901}. For a composite system $\rho_{AB}$, the quantum discord is written in the following form 
\begin{align}
\mathcal{D}_{\text{OZ}}(\rho)=\mathcal{I}(\rho)-\mathcal{J}(\rho)	 ,
\end{align}
where $\mathcal{I}(\rho)=S(\rho_{A})+S(\rho_{B})-S(\rho_{AB})$ determines the mutual correlation of the composite system, and $\mathcal{J}(\rho)=S(\rho_{B})-\min_{\lbrace\Pi_{k}^{A}\rbrace}\sum_{k}p_{k}S(\rho_{B\vert k})$ quantifies the classical correlation with the optimization over all possible POVMs ${\lbrace\Pi_{k}^{A}\rbrace}$, defined in the Hilbert space of the subsystem $A$ with $p_{k}=\mathrm{tr}(\Pi_{k}^{A}\rho)$ and $\rho_{B\vert k}=\mathrm{tr}_{A}(\Pi_{k}^{A}\rho)/p_{k}$.

Given that the quantum discord calculation is a complicated optimization process, many efforts have been made to provide a complete treatment for the quantum discord of two-qubit in the particular case where the system is in a X-state~\cite{PhysRevA.81.042105,PhysRevA.83.012327,PhysRevA.84.042313}. Therefore, it is worth to investigate our results through the analyzes introduced in Ref.~\cite{PhysRevA.84.042313}. The authors identify the set of optimal measurements to compute the discord of a X-state. By using the theorem presented in Re.~\cite{PhysRevA.84.042313}, it is possible to show that the optimal measurement for quantum discord of Werner state $\rho_{\mathrm{w}}$ is given by $\sigma_{z}^{A}$. By doing that, one obtains
\begin{align}\label{A2}
\mathcal{D}_{\text{OZ}}(\rho_{\mathrm{w}})&=\frac{1-\varepsilon}{4}\ln(1-\varepsilon)+\frac{1+3\varepsilon}{4}\ln(1+3\varepsilon)\nonumber\\
&-\frac{1+\varepsilon}{2}\ln(1+\varepsilon)	 .
\end{align}
As an important result obtained in this approach, it is straightforward to display $\mathcal{D}_{\text{OZ}}(\varrho_{\rho_{\mathrm{w}}}^{H})$ in the regime where $\omega< J/\sqrt{2}$ is equal to $\mathcal{D}_{\text{OZ}}(\rho_{\mathrm{w}})$ in Eq.~\eqref{A2}. This meets a good agreement with the previous result highlighted in the paper, where one can extract ergotropy without change the quantum correlations of the system. Note that other passive states obtained in the main text are classical states, than they present quantum discord equals to zero. Then, in summary, by comparing the above result with the Eqs.~\eqref{ergo} and~\eqref{ergoH}, of the main text, it is possible to conclude that the trace distance discord is more efficient than the Olliver-Zurek discord in describing the relation between ergotropy and quantum correlations.

\bibliography{mybib-URL.bib}
	
	
\end{document}